\documentstyle[12pt]{article}
\begin{document}
\pagestyle{empty}
\nonumber
{\it Invited talk at International Symposium
on Exotic Atoms and Nuclei, Hakone, Japan, June 7--10, 1995}
\begin{center}
\begin{large}
{\bf Energy--level splitting in antiprotonic helium atoms}\\
\vspace{0.3truecm}
O.~I.~Kartavtsev\footnote{{\it e-mail address:}
oik@thsun1.jinr.dubna.su}\\[0.3cm]
\end{large}
Bogoliubov Laboratory of Theoretical Physics \\
Joint Institute for Nuclear Research\\
141980, Dubna, Russia\\
Fax: 7 096 21 65084\\
\end{center}

\sloppy

\begin{abstract}
Recent experiments
on the laser-induced resonant annihilation provide a precise measurement
of transition energies of antiprotonic helium atoms.
Relativistic corrections of an
order of $\alpha ^2$ to the pure Coulomb
interaction will be taken into account in the theoretical description
of energy spectra of antiprotonic helium atoms. The splitting
of energy levels arising due to the spin-dependent part of the
relativistic interaction is considered for a number of states of
$^{3,4}\!  H\!  e\bar pe$ systems.

\end{abstract}
\newpage
\section{Introduction}

Metastable antiprotonic helium atoms $^{3,4}\! He\bar pe$ have been
discovered in experiments on the delayed annihilation of
antiprotons in helium media~\cite{iwa}, \cite{nak}.
Analogous long-lived systems were observed in
experiments with negative kaons~\cite{yam} and pions~\cite{nakpi}.
The discussion of the theoretical calculations on antiprotonic helium
atoms and related topics can be found in~\cite{fb}.

The precise measurement
of transition energies of antiprotonic helium atoms in recent
experiments on the laser-induced resonant annihilation~\cite{mor},
{}~\cite{hay},~\cite{maas} invokes the theoretical description of
energy spectra with comparable accuracy. Such a description of energy
spectra requires that minor effect of
relativistic and QED interactions and the coupling with the
continuous spectrum should be taken into account.  The
relativistic corrections of an order of $\alpha ^2$ ($\alpha
=e^2/\hbar c$ -- fine structure constant) to the pure Coulomb
interaction are the most important and should be firstly considered.
Next in line are QED corrections to energies of higher orders in
$\alpha $.

Since the contribution to energies from the relativistic interactions
depends on the antiproton mass, charge and magnetic moment, the comparison of
precise calculations and measurements of the energy spectra can be
used for determining the antiproton properties.
This knowledge is significant in testing the  fundamental symmetry
principles. The detailed discussion of this problem can be found
in~\cite{eades}.

The spin-dependent part of the relativistic interactions gives rise
to splitting of energy levels, and each single transition turns into
a multiplet. Sufficiently large distances between lines in the
multiplet can be measured experimentally. It is worthwhile to
mention that the resolution in current experiments is about 10GHz
and without much difficulty can be improved to 1GHz~\cite{pps}.
As it will be discussed below, due to the interaction with electron
spin, antiprotonic helium energy levels split into two multiplets
and the interaction with nuclei spins provides a minor splitting
into each multiplet. Values of the former large splitting are
presented in this report.

The relativistic interaction in antiprotonic helium atoms
is described in the next section, and in section~3 the method of
calculation of the level splitting is discussed. Numerical results
are presented in section~4; outlook and discussion, in the last
section.

\section{Relativistic interaction}

For each pair of particles $i, j$ in the three-body system the
relativistic correction of an order of $\alpha ^2$ to the pure
Coulomb two-body potential can be described by the Breit interaction
of the form

\begin{eqnarray}
U_{ij}=-\alpha ^2\! Z_iZ_j\Bigl(\displaystyle\frac{\pi}{2}(m_i^{-2}+m_j^{-2})
\delta({\bf r}_{ij})+\frac{1}{2m_im_jr_{ij}}\bigl({\bf p}_i{\bf p}_j+
r_{ij}^{-2}
{\bf r}_{ij}({\bf r}_{ij}{\bf p}_i){\bf p}_j\bigr)+\nonumber \\
\displaystyle\frac{\mu_i}{2m_i^2r_{ij}^3}
[{\bf r}_{ij}{\bf p}_i]{\bf s}_i-
\frac{\mu_j}{2m_j^2r_{ij}^3}[{\bf r}_{ij}{\bf p}_j]{\bf s}_j
+\frac{1}{m_im_jr_{ij}^3}(\mu_j[{\bf r}_{ij}{\bf p}_i]{\bf s}_j-
\mu_i[{\bf r}_{ij}{\bf p}_j]{\bf s}_i)-\\
\displaystyle\frac{\mu_j\mu_j}{m_im_j}
\bigl(r_{ij}^{-3}{\bf s}_i{\bf s}_j-3({\bf r}_{ij}{\bf s}_i)({\bf r}_{ij}{\bf
s}_j)-
\frac{8\pi}{3}{\bf s}_i{\bf s}_j\delta({\bf r}_{ij})\bigr)\Bigr),\nonumber
\label{breit}
\end{eqnarray}
where ${\bf r}_{ij}={\bf r}_{i}-{\bf r}_{j}$,
${\bf r}_{i}, {\bf p}_{i}, {\bf s}_{i}, m_i, Z_{i}, \mu_i$
are the radius-vector, momentum, spin, mass, charge and magnetic
moment (in units $e\hbar /2m_ic$) of particle $i$.
Here and below atomic units ($\hbar =e=m_e=1$) are used.
The correction to
the kinetic energy of an order of $\alpha ^2$ for each particle $i$
is
\begin{equation}
\Delta T_{i}=-\frac{\alpha
^2}{8}\frac{p_i^4}{m_i^3} \label{kin}
\end{equation}
Full relativistic correction $H_r$ of an order of $\alpha ^2$ to the
three-body nonrelativistic Hamiltonian is a sum of $U_{ij}$ for all
pairs of particles and $\Delta T_i$ for all particles
\begin{equation}
H_{r}=\sum_{i}\Delta T_i+\sum_{i>j}U_{ij}.
\label{relint}
\end{equation}
Using expressions $U_{ij}, \Delta T_i$  in eq.~(\ref{relint}),
particles momenta ${\bf p}_i$ will be taken in the center of mass
frame of the three-body system~\cite{lev}.  Below, for definiteness
 helium nucleus, antiproton and electron will be enumerated as
 particles $1,2,3$, respectively. Relative
 coordinates ${\bf r}={\bf r}_{2}-{\bf r}_{1}$,
$\mbox{\boldmath$\rho$}={\bf r}_{3}-{\bf r}_{1}$, corresponding momenta
${\bf p}=-i\nabla_{\bf r}$, \
${\bf q}=-i\nabla_{\mbox{\boldmath$\rho$}}$
and angular momenta
${\bf l}=[{\bf rp}]$, \
$\mbox{\boldmath{$\lambda$}}=[\mbox{\boldmath$\rho$} {\bf q}]$
will be used to simplify notation.

The interaction  $H_r$, given in (\ref{relint}), conserves the sum
${\bf J}={\bf L}+\sum_{i}{\bf s}_i$ of
 the total angular momentum $\bf L={\bf l}+\mbox{\boldmath$\lambda$}$
and particle spins ${\bf s}_i$. Each level of the
nonrelativistic Hamiltonian splits into four and eight sublevels
for $^{4}\! H\! e\bar pe$ and $^{3}\! H\! e\bar pe$ systems, respectively.
Due to
very small mass ratios $m_3/m_1, m_3/m_2$, the largest contribution to
the energy splitting comes from the interaction with the electron
spin ${\bf s}_3$.
Taking into consideration only terms responsible for the splitting
in (\ref{breit}), (\ref{relint}),
this part of relativistic interaction can be written as follows:
\begin{eqnarray}
H_s=\alpha^2\bigl(\frac{1}{\rho^{3}}\mbox{\boldmath$\lambda$}{\bf s}_3+
\frac{1}{2|{\bf r}-\mbox{\boldmath$\rho$}|^3}
([{\bf r}-\mbox{\boldmath$\rho$},{\bf q}]{\bf s}_3)-
\\
\frac{1}{m_2|{\bf r}-\mbox{\boldmath$\rho$}|^3}
([{\bf r}-\mbox{\boldmath$\rho$},{\bf p}]{\bf s}_3)+
\frac{2}{m_1\rho^3}([\mbox{\boldmath$\rho$},{\bf p}]{\bf s}_3)\bigr)\nonumber
\label{split}
\end{eqnarray}
While the last two terms in (\ref{split})
are inversely proportional to the heavy particle masses $m_{1,2}$,
their contribution to the energy splitting is nevertheless comparable
to the contribution from the first two terms due to the following
reasons.  The small mass factor is compensated in part due to the
large angular momentum $l \sim L$ of heavy particles.  At the same
time, only small components of the wave function corresponding to the
nonzero electron angular momenta $\lambda \ne 0$ lead to the
nonzero splitting value from the first two terms in (\ref{split}).

\section{Level splitting}

The interaction  $H_s$, given in (\ref{split}), conserves the sum
${\bf j}={\bf L}+{\bf s}_3$ of
 the total angular momentum $\bf L={\bf l}+\mbox{\boldmath$\lambda$}$
and electron spin ${\bf s}_3$ and
splits each level into two sublevels, corresponding to the
eigenvalues $j=L\pm 1/2$.
The part of the interaction depending on heavy particle
spins removes the remaining degeneracy and splits each $j=L\pm 1/2$
sublevel further into two or four levels for the $^{4}\! H\! e\bar
pe$ and $^{3}\! H\! e\bar pe$ systems, respectively. Values of
this secondary splitting are much smaller in comparison with the
splitting, arisen due to the interaction with the electron
spin~(\ref{split}). By this reason only calculation of major
splitting will be presented in this report.

The nonrelativistic Hamiltonian of the antiprotonic helium atom is
\begin{equation}
H=-\frac{1}{2\mu}\Delta _{\bf r}-\frac{1}{2\mu_{1}}\Delta
_{\mbox{\boldmath$\rho$} }
-\frac{1}{m_1}{{\bf\nabla_r}\cdot{\bf\nabla}_{\mbox{\boldmath$\rho$} }}
-\frac{2}{r}-\frac{2}{\rho }+\frac{1}{|{\bf r}-\mbox{\boldmath$\rho$} |},
\label{ham}
\end{equation}
where $1/\mu=1/m_1+1/m_2,~1/\mu_1=1/m_1+1/m_3$.
The nonrelativistic wave function $\psi_{LN}$ and energy $E_{LN}$ is
the $N$-th solution and eigen-energy of the Schr\"{o}dinger equation
\begin{equation}
(H-E_{LN})\psi_{LN}=0
\label{schr}
\end{equation}
for the total angular momentum $L$.
Since the splitting is small in comparison with energy
differences between states of different $L$ values, the energy
shift $\Delta_{LN}$ can be found in the first order of perturbation
theory over $H_{s}$
\begin{equation}
\Delta_{jLN}=\langle \Psi_{jLN}|H_s|\Psi_{jLN}\rangle,
\label{shift}
\end{equation}
where $\Psi_{jLN}$ is the vector production of $\psi_{LN}$ and spin
function describing the dependence of the electron spin.

Since the interaction $H_s$~(\ref{split}) is of the form
$H_s={\bf As}_3$, the energy shift $\Delta_{jLN}$ can be expressed
\begin{equation}
\Delta_{jLN}=
\frac{j(j+1)-L(L+1)-3/4}{2\sqrt{L(L+1)(2L+1)}}
\langle \psi_{LN}||{\bf A}||\psi_{LN}\rangle,
\label{shift1}
\end{equation}
where the notation $\langle \cdot ||\cdot ||\cdot \rangle$ stands for
the reduced matrix element.
Level splitting
$\Delta E_{LN}=\Delta_{L+1/2LN}-\Delta_{L-1/2LN}$
is a difference of shifts~(\ref{shift1}) for
the $j=L\pm 1/2$.

Due to smallness of the relativistic interaction, radiative
transitions proceed only between states of the same $j$. By this
reason, in experiment each spectral line of the transition
from the state $L_iN_i$ to state $L_fN_f$ is to be split into a
doublet with the interline distance $\Delta\nu =\Delta
E_{L_iN_i}-\Delta E_{L_fN_f}$.

\section{Numerical results}

The variational method, described in~\cite{kar}, was
applied to determine eigenfunctions and eigenenergies of the
Schr\"{o}dinger equation (\ref{schr}).  The set of simple
variational trial functions of the form
\begin{equation}
\chi_{nkl\lambda i}^{LM}({\bf r}, \mbox{\boldmath$\rho$})=
{\cal Y}_{l\lambda}^{LM}({\bf\hat r},{\bf\hat{\mbox{\boldmath$\rho$}}})
r^{l+i}\rho^{\lambda} exp(-a_nr-b_k\rho ) ,
\label{trial}
\end{equation}
where ${\cal Y}_{l\lambda}^{LM}({\bf\hat r},{\bf\hat{\mbox{\boldmath$\rho$}}})$
are bispherical harmonics of angular variables,
was used in the calculations.

Splitting values $\Delta E_{LN}$
have been calculated as described above (\ref{shift}), (\ref{shift1})
by using variational nonrelativistic wave functions.
Up to 600 trial functions~(ref{trial}) containing up to 15
bispherical harmonics were used in these calculations. Nonlinear
parameters $a_n, b_k$ were taken the same as in the previous
variational calculation of energies and radiative transition
rates~\cite{kar}.

Splitting values for a number of states of the $^{3,4}\! He\bar pe$
systems in the range of experimentally observed values of the total
angular momentum $L$ are presented in Table 1.

Table~1. Splitting values $\Delta E_{LN}$ ($10^{-6}$au) of
the lowest levels in the $^{3,4}\! H\! e\bar pe$ systems.

\noindent
\begin{tabular}{lllllll} \hline\hline
\multicolumn{7}{c}{$^{4}\! H\! e\bar pe$}\\ \hline
N & L=32 & L=33 & L=34  & L=35  & L=36  &  L=37 \\ \hline
1 & -1.10 & -1.15 & -1.15 & -1.14 & -1.12 & -1.09 \\ \hline
2 & -1.12 & -1.09 & -1.08 & -1.07 & -1.04 & -1.00 \\ \hline
3 & -1.01 & -1.02 & -1.00 & -0.98 & -0.94 & -0.90 \\ \hline
4 &       & -0.94 & -0.94 & -0.90 & -0.86 & -0.82 \\ \hline
5 &       &       & -0.93 & -0.90 & -0.84 & -0.81 \\ \hline   \hline
\multicolumn{7}{c}{$^{3}\! H\! e\bar pe$}\\ \hline
N & L=31  & L=32  & L=33  & L=34  & L=35  & L=36 \\ \hline
1 & -1.20 & -1.16 & -1.19 & -1.19 & -1.18 & -1.14 \\ \hline
2 & -1.14 & -1.19 & -1.15 & -1.12 & -1.08 & -1.04 \\ \hline
3 & -1.08 & -1.06 & -1.05 & -1.04 & -1.00 & -0.98 \\ \hline
4 &       & -0.97 & -0.92 & -0.86 & -0.85 & -0.81 \\ \hline
\hline
\end{tabular} \\

As it follows from expression~(\ref{split}), the form of the wave
function at small interparticle distances is the most important
 in evaluating the integral~(\ref{shift}). Convergence of the
calculated splitting provides a few per cent
relative accuracy. It is worthwile to mention
that due to the variational method of calculation the accuracy
is better for the large $L$ and small $N$ states.  It is impossible
to trace the convergence in the case of short lived states due
to a small multipolarity $\Delta l<3$ of the Auger decay. This
problem is closely connected with a large natural width of these
states which exceeds significantly a splitting value.  Also, the
variational procedure meets some difficulties in describing the short
range behavior of the wave function for the large enough $N$ states,
especially in the $^{3}\! H\! e\bar pe$ system. These are reasons to
omit the above--mentioned cases in Table~1.

The last two terms in eq.~(\ref{split}) describe the interaction of the
electron magnetic moment with the magnetic field of heavy particles.
These terms give rise to the largest contribution to the energy--level
splitting. For better understanding the splitting dependence on $L,N$
this contribution is presented in Table~2 for the $^{4}\! H\! e\bar
pe$ system. The contribution to the energy--level splitting from the
first two terms in~(\ref{split}) are of the opposite sign and much
smaller in magnitude. Nevertheless, decreasing in this contribution with
increasing $L$ compensates the $L$ dependence of the last two terms in
eq.~(\ref{split}) and provides a very slow dependence of the total
splitting $\Delta E_{LN}$ on $L$.

Table~2. Contribution of the last two terms in eq.~(\ref{split})
to the energy--level splitting $\Delta E_{LN}$ ($10^{-6}$au) in the
$^{4}\! H\! e\bar pe$ system.

\noindent
\begin{tabular}{lllllll} \hline\hline
N & L=32  & L=33  & L=34  & L=35  & L=36  &  L=37 \\ \hline
1 & -1.41 & -1.43 & -1.40 & -1.37 & -1.34 & -1.28 \\ \hline
2 & -1.39 & -1.34 & -1.30 & -1.27 & -1.22 & -1.16 \\ \hline
3 & -1.26 & -1.24 & -1.20 & -1.15 & -1.10 & -1.04 \\ \hline
4 &       & -1.14 & -1.11 & -1.06 & -1.00 & -0.94 \\ \hline
5 &       &       & -1.10 & -1.06 & -0.98 & -0.95 \\ \hline   \hline
\end{tabular} \\

\section{Discussion}

Due to almost exact conservation of the $j$ value in the radiative
transition the spectral line splitting will be found as a difference
of $\Delta E_{LN}$ presented in Table~1.
Most appropriate for the experimental measurement are
the favoured transitions between states of the same
$N$, which have the largest
radiative rates~\cite{kar}--\cite{shim}. However, the
calculated splitting values are almost independent of $L$ for a
given $N$, and it is not plausible to resolve such a small
difference in splitting for the favoured transitions.  For this
reason, the experimental proposal for the near future~\cite{pps} is
aimed at searching for the splitting in unfavoured  transitions
$(L,N) \to (L-1,N+2)$.

In order to measure splitting in experiments on the laser--induced
resonant annihilation the initial state will be long--lived. This is
provided by the condition that the multipolarity of the Auger decay for
this state is $\Delta l=4$. The next condition is that the natural
width of the short--lived final state will be smaller than the splitting
value, and the multipolarity of the Auger decay for this state will be
$\Delta l=3$.
The spectral line splitting for a number of suitable transitions
is presented in Table~3. These values are of an order
of the experimentally measurable value $\sim $1GHz.

Comparing the splitting values for the $^{4}\! H\! e\bar pe$ and
$^{3}\!  H\! e\bar pe$ systems one can mention in the
$^{4}\! H\! e\bar pe$ case
a slower decreasing in $\Delta E_{LN}$ with increasing $N$. As it is
clear from Table~3, this isotopic effect is also conserved for the
spectral line splitting $\Delta\nu$.

Table~3. Spectral line splitting
$\Delta\nu =\Delta E_{L_iN_i}-\Delta E_{L_fN_f}$ (GHz) for the
transitions $E_{L_iN_i} \rightarrow E_{L_fN_f}$
in the $^{3,4}\! H\! e\bar pe$ systems.

\noindent
\begin{tabular}{r@{$\rightarrow$}lc|r@{$\rightarrow$}lc} \hline\hline
\multicolumn{3}{c|}{$^{4}\! H\! e\bar pe$} &
\multicolumn{3}{c}{$^{3}\! H\! e\bar pe$}\\ \hline
$L_iN_i$ & $L_fN_f$ & $\Delta\nu$ & $L_iN_i$ & $L_fN_f$ & $\Delta\nu$
\\ \hline
33,1 & 32,3 & -0.92 & 32,1  & 31,3  & -0.53 \\ \hline
34,1 & 33,3 & -0.86 & 33,1  & 32,3  & -0.86 \\ \hline
34,2 & 33,4 & -0.91 & 33,2  & 32,4  & -1.22 \\ \hline
35,2 & 34,4 & -0.88 & 34,2  & 33,4  & -1.35 \\ \hline
35,3 & 34,5 & -0.34 & \multicolumn{3}{c}{}  \\ \hline
\hline
\end{tabular} \\

The following considerations can be used to understand
qualitatively the $L,N$--dependence of the energy--level
splitting. Contribution to splitting from
the interaction of the electron magnetic
moment with the magnetic field of heavy particles is described by
the last two terms in the splitting interaction $H_s$~(\ref{split}).
This contribution is
proportional to the relative momentum of heavy particles $\bf p$.One
can consider that the motion of heavy particles is approximately the
same as in a hydrogen--like atom and momentum $p$ is inversely
proportional to the angular momentum $L$. This is the reason for
increasing this contribution with decreasing $L$, as presented in Table~2.
The contribution from the first two terms in the splitting interaction
$H_s$ is
connected with the electron rotation and proportional to the small
component of the wave function arising due to polarization of an
electron by $\bar p$.
With decreasing $L$ the antiproton moves to a region of increasing
electron density and the polarization increases. In such a way
contributions to the energy--level splitting from
the last two terms in $H_s$ and remaining part of splitting interaction
are of opposite signs and level
off the dependence of the total splitting $\Delta E_{LN}$ on $L$.

One can consider quasiclassically that the antiproton orbit
became more stretched with increasing $N$ at fixed
total angular momentum. By this reason all the terms of the
splitting interaction $H_s$ decrease with encreasing $L$ and provide
the $N$ dependence presented in Tables~1,~3.

{\it Acknowledgement.}
Fruitful discussions with Dr.~J.~Eades
and Prof.~T.~Yamazaki were important in the initiation of this
work.
The author would like to thank Computing and Network
Division of CERN for the valuable assistance
in computing necessary for these calculations.
The author is grateful to the Japanese Society for
Promotion of Science for financial support of the participation at
the International Symposium on Exotic Atoms and Nuclei.

\end{document}